# Towards a compact all optical terahertz-driven electron source at Tsinghua University

Hanxun Xu[1,2], Renkai Li[1,2], Lixin Yan[1,2], Yingchao Du[1,2], Qili Tian[1,2], Wenhui Huang[1,2*] and Chuanxiang Tang[1,2]

[1] *Key Laboratory of Particle and Radiation Imaging, Tsinghua University, Ministry of Education, Beijing CN-100084, China.*
[2] *Department of Engineering Physics, Tsinghua University, Beijing CN-100084, China.*
*\*huangwh@mail.tsinghua.edu.cn*

**Abstract:** We propose a physical design of a compact all optical terahertz (THz)-driven electron source. The 300 mm accelerator beamline, powered by Joule level laser system, is easily to be integrated to tabletop scale. A dual-feed THz-driven electron gun with an exponential impedance, a tapered dielectric loaded cylindrical waveguide, THz-driven bunch compressors and permanent magnet solenoids (PMS) have been designed and optimized. Dynamics simulations show that the electron source can deliver a 19 fC, 3 MeV electron beams with a normalized transverse emittance of 0.079 π.mm.mrad. A minimum relative energy spread of 0.04% or a minimum root-mean-square bunch length of 6.1 fs can be achieved by adjusting the beam shaping line. Sensitivity analysis shows that the THz-driven electron source can effectively work under a 1.5% energy jitter of the THz power system. Simulated diffraction pattern up to the fourth order of an aluminum sample based on the beamline can be clearly distinguished. A prototype THz gun has beam fabricated and is now under testing, more results will be reported in future works.

## 1. Introduction

Conventional particle accelerators are powerful tools for the development of modern science and technology. Owing to the radio-frequency induced plasma breakdown [1-3], the acceleration gradients of microwave structures are difficult to be improved, resulting in large and expensive accelerator facilities. Novel acceleration techniques have been proposed to produce and sustain higher acceleration gradients, thus, shrink the accelerator size and cost. One promising technique is to up-convert the work frequency of the accelerator from microwave to terahertz wave [4-6] or optical regime [7-9], since higher frequency and shorter pulse length contributes to higher breakdown threshold [1]. Plasma wakefield accelerations [10-12] are also very attractive by using intense laser pulse or high density electron bunch to produce strong wakefield for accelerating particles. Recently, acceleration and manipulation of high quality electron beams using THz wave have shown their great potential for achieving tabletop accelerators. THz-driven acceleration offers GV/m acceleration gradient and pC level bunch charge, providing a good compromise between conventional accleration and direct laser acceleration. The millimeter scale structures also relieve the manufacturing difficulties and timing control tolerances. Great progress has been made in THz-based electron guns [4,13], linear acceleration [5-6,14] staging [15-17], streaking [18-19], compression and timing jitter suppression [20-21], making it feasible to build real THz-driven accelerators [22-24].

In this paper, we report the physical design of a compact all optical THz-driven electron source based on our previous works [16, 24-26] at Tsinghua University. The general design of the accelerator beamline and the required power system will be proposed. A dual-feed THz-driven electron gun with an exponential impedance has been designed and fabricated to produce a 20 pC, 55 keV electron beam. A tapered dielectric loaded cylindrical waveguide has been developed to boost the beam to 3 MeV. Beam shaping component including two THz-driven compressors and three permanent solenoid magnets will also be presented. Dynamics simulations are carried out to optimize the transvers emittance of the injector to 0.079 π.mm.mrad, and minimize the relative energy spread or the root-mean-square bunch length at the end of the beamline to 0.04% or 6.1 fs for different user requirements. Sensitivity analysis has been carried out to determine the impact of the energy jitter on the beam parameters. Simulation of the diffraction pattern of the aluminum sample using the generated electron beam also shows that the designed THz-driven electron source holds great potential for achieving high quality MeV ultrafast electron diffraction (UED) facilities. The current status of this project will also presented in this paper.

## 2. General design

### 2.1 Accelerator beamline

Fig. 1 shows the schematic of the compact all optical THz-driven electron source. The accelerator beamline consists of an injector followed by a beam shaping line. The injector includes a 55 keV THz-driven electron gun, a rectangular waveguide bunch compressor, a THz linac and a PMS for emittance compensation. A 30 fs (FWHM) ultraviolet laser illuminate the copper cathode of the THz gun, generating the photo electron bunch. Two 0.45 THz, 108 μJ single cycle THz pulse drive the dual-feed THz gun, exciting the TE01 mode electromagnetic field which accelerates the photo electron bunch from rest to 55 keV. The compressor downstream the gun compresses the beam with two 0.45 THz, 2 μJ single cycle THz pulse. A tapper dielectric loaded waveguide (DLW), powered by a 0.45 THz, 280 ps (FWHM), 2.64 mJ THz wave, boosts the compressed beam to 3MeV. A PMS enclosing the DLW has been designed to optimize the transverse emittance at the exit of the injector. The beam shaping line consists of two PMSs and a DLW compressor to manipulate the phase space of the electron beam at the end of the beamline. PMS2 focuses the beam from the injector to the DLW compressor. The DLW compressor works at zero-crossing phase to compress the bunch length or minimize the energy spread. The DLW can also work at maximum accelerating phase as a post acceleration stage to further increase the beam energy. The last PMS3 refocuses the electron beam at the user target point. The total length of the beam line is about 300 mm while the length of the injector is only 25 mm. This compact THz-driven accelerator beamline can be easily integrated to tabletop size.

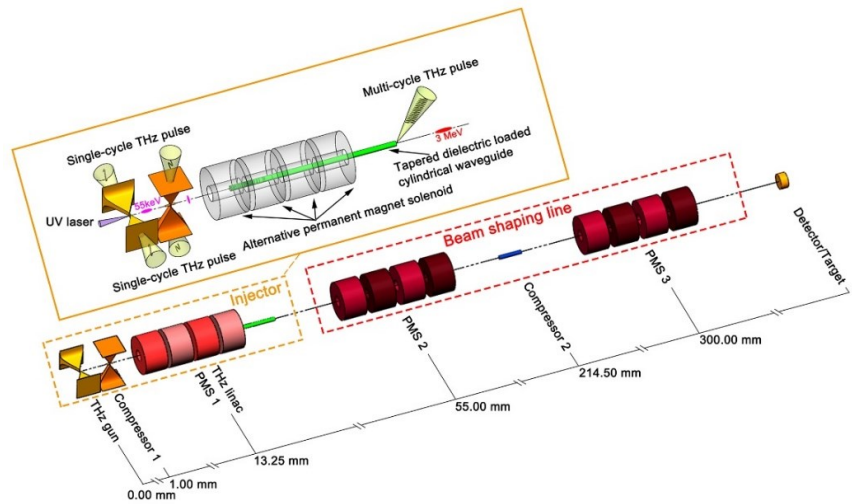

Fig. 1. The schematic of the compact all optical THz-driven electron source.

### 2.2 Power system

The THz-driven electron beam line works at 0.45 THz and is powered by a Joule level infrared laser system, as shown in Fig. 2. The photocathode driven laser can be generate from infrared laser via optical third harmonic generation method. The single cycle THz pulses driving the THz gun and THz compressor 1 are produced by optical rectification with a laser-

THz conversion efficiency about 1% [26-27]. Optical generation of sub-mJ level narrowband THz radiation based on periodically-poled lithium niobate (PPLN) has been demonstrated [28], this approach can be further developed to generate the mJ level multi-cycle THz pulse that required to drive the THz linac. The few-cycle THz pulse driving the DLW compressor 2 can be produced by combining the optical rectification approach with a pulse stacking method [26]. All the THz wave and the cathode driven laser are generated from the same seeding laser, which benefits the inherent synchronization between the THz field and the electron bunch. One can also pump the sample with a split infrared laser and then probe the dynamic process using the generated electron bunch, this pump-probe setup enables precise temporal resolution supported by the inherent synchronization between the pump laser and the probe electron beam.

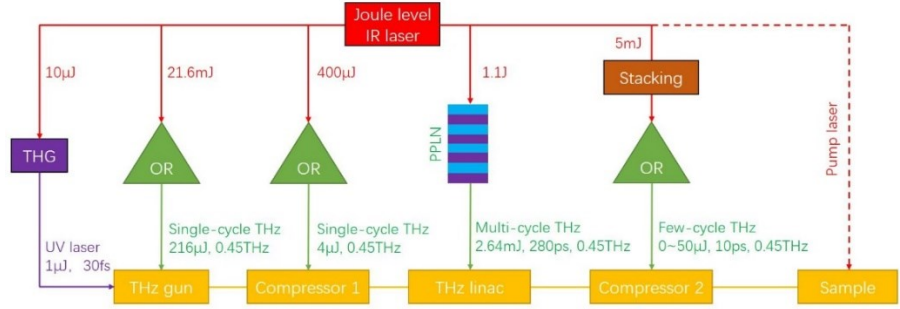

Fig. 2. The power system of the compact all optical THz-driven electron source.

## 3. Beamline components

### 3.1 THz gun

The dual-feed THz-driven electron gun is a tapered rectangular waveguide with an exponential impedance along the wave propagation direction as shown in Fig. 3a. The tapered rectangular waveguide can focus the electric field and magnetic field simultaneously. The dual-feed work scheme enables stronger acceleration field while eliminating the magnetic deflecting force, since the electric field doubling and the magnetic field cancellation appears in the gun center. An exponential impedance variation benefit a higher coupling efficiency and broaden coupling bandwidth. Assuming a Cartesian coordinates whose origin lies in the gun center, x/z axis parallel with the waveguide width/height direction, and y axis is the THz propagation direction. The wave impedance of the TE10 mode in a rectangular waveguide is given as:

$$\eta_{TE10} = \frac{\eta_0}{\sqrt{1-\left(\frac{\lambda}{2a}\right)^2}} \quad (1)$$

Where $\eta_0$ is the wave impedance of the free space and $\lambda$ is working wavelength. We taper the waveguide width to fulfill the exponential impedance variation, resulting in a variation of the waveguide width as:

$$a(y) = \frac{\lambda}{2\sqrt{1-\left(\frac{\eta_0}{A\exp(B(L-y))}\right)}} \quad (2)$$

Where A and B are coefficients determined by the waveguide dimension, L is the half length of the gun. The gun locate in the focus spot of the THz wave, where the THz beam could be described by a gauss beam. To further increase the coupling efficiency, the waveguide height also varies in the similar form as the beam waist of the gauss beam:

$$b(y) = b_0\sqrt{1+\left(\frac{y-L_1}{k}\right)^2} \quad (3)$$

Where $b_0$ and $k$ are coefficients determined by the waveguide dimension, $L_1$ is the half length of the uniform waveguide at the gun center. CST [29] simulations have been carried out to optimize the coupling efficiency and the field distribution of the gun powered by a single cycle THz wave reported in ref [27]. Fig. 3 shows the distribution of the acceleration field in the cross sections of the gun. The acceleration field in the gun center has been optimized to be homogeneous and symmetric, which will contribute to a smaller transverse emittance growth. A field enhancement of 3.5 takes place at the gun center.

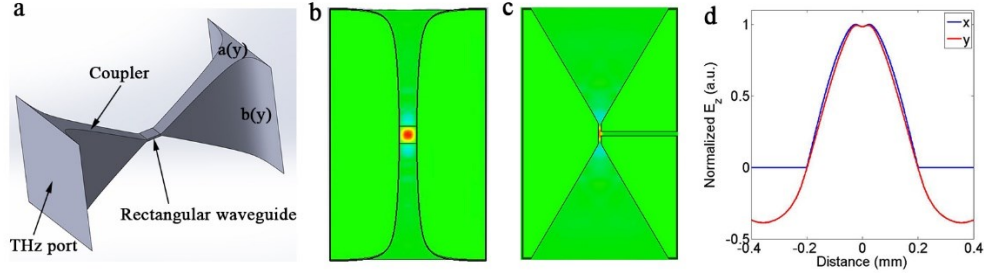

Fig. 3. The dual-feed THz-driven electron gun. (a) the vacuum structure of the gun, (b) the distribution of the acceleration field in the x-y plane, (c) the distribution of the acceleration field in the z-y plane. (d) the normalized acceleration field amplitude along x and y axis..

Fig. 4 shows the dynamic simulation results of the THz gun using CST PIC studio. The initial electron beam is generated via ASTRA [30] code. The temporal profile of the driven laser is 30 fs FWHM Gaussian distribution. According to our laser parameters, the transverse intensity distribution is homogeneous with a radius of 25 μm. Powered by two 108 μJ single cycle THz pulse, the gun successfully accelerates the initial beam from rest to 55 keV with an r.m.s. energy spread of 0.8 keV. The normalized transverse emittance at the gun exit is 0.013 π.mm.mrad with a bunch charge of 20 fC. Owing to the space charge effect and the energy spread induced speed divergence, the bunch length grow linearly after emits from the gun exit.

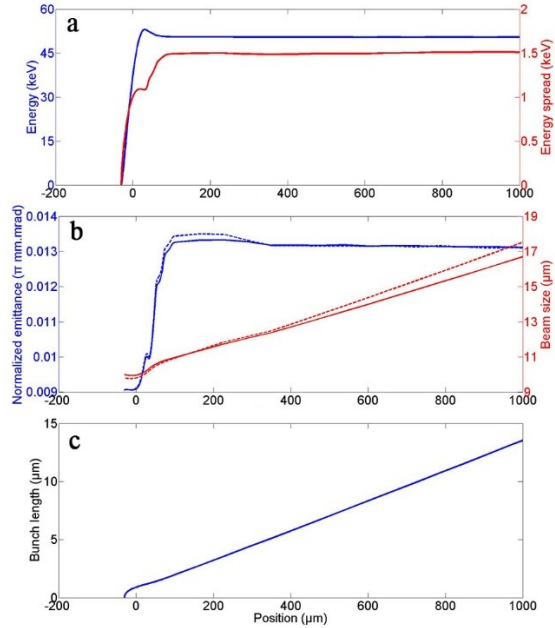

Fig. 4. Dynamics simulation results of the THz gun. (a) mean energy and r.m.s. energy spread, (b) normalized transverse emittance and r.m.s. beam size, solid/dotted lines are parameters in the x/y direction, (c) r.m.s. bunch length

### 3.2 Main linac

The main linac is a 16.5 mm tapered DLW with a constant inner diameter of 400 μm. The speed of the nonrelativistic electron beam increase significantly when they are accelerated by the THz field. We tapered the dielectric ($Al_2O_3$, $\varepsilon_r = 10.56$) thickness of the DLW so that the phase velocity of the TM01 mode traveling field matches the speed of the electron beam. Lemery, F. et. al. have derived an analytical model to describe the field evolution and the beam dynamics of a tapered DLW [31]. Other than using the analytical model, we use CST MWS studio to calculate the field amplitude, the phase velocity and the wave propagation constant of different DLWs, and then calculate the single particle dynamics of the DLW using a fourth-order Runge-Kutta algorithm to solver the kinematic equation:

$$\frac{\partial z}{\partial t} = \beta c \quad (4)$$

$$\frac{\partial \beta}{\partial t} = \frac{eE(z)}{\gamma^3 mc} \sin\left(\int_0^z k_z(z)dz + \omega t + \varphi_0\right)$$

Where $\beta$ is the normalized speed of the electron, $\varphi_0$ is the phase constant, $\omega$ is the working frequency. The field amplitude $E(z)$ and the propagation constant $k_z(z)$ is given by CST calculation. Fig. 5 shows the simulation results of the designed tapered DLW. The field amplitude increase as the dielectric thickness decrease. The main linac boosts the beam from the THz gun exit to 3 MeV with a 0.45 THz, 2.64 mJ, 280 ps narrowband THz wave. The effective acceleration gradient of the tapered DLW is 182 MV/m with a peak field of 285 MV/m.

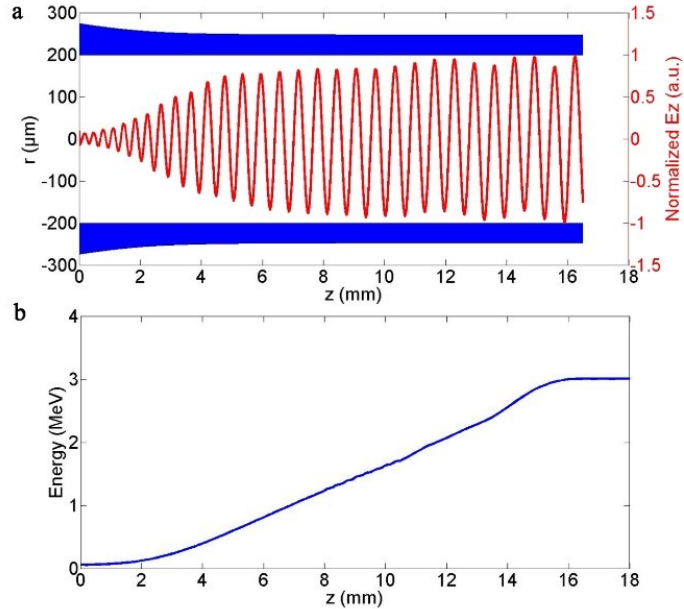

Fig. 5. The tapered DLW. (a) the dielectric dimension and the normalized field amplitude, (b) the energy variation of the single particle simulation.

### 3.3 Bunch compressor

The acceleration phase bandwidth taken by the electron need to be as narrow as possible in order to achieving higher acceleration efficiency. The wavelength of the THz wave at the main linac entrance/exit is 0.29/0.67 mm. For a 10° phase bandwidth, the correspond beam length is 8/18.5 μm. The bunch length grows quickly after emit from the THz as shown in Fig. 4c. A THz bunch compressor is located after the gun to compress the bunch length growth so that it can fulfill the length requirement of the main linac. The compressor shares the waveguide structure as the THz gun as shown in Fig. 6a. The main difference is that the compressor works at a zero-cross phase where velocity bunching takes place. Owing to the speed divergence and energy spread, the beam emitted from the gun has a negative energy chirp, the electron energy at the bunch head is high than that at the bunch tail. When injected at the negative gradient zero-cross phase, the bunch head is decelerated while the bunch tail is accelerated. The energy chirp will be flipped under certain THz field strength. The beam energy at the beam head is lower than that at the bunch tail after the beam leave the bunch compressor. The electrons at the bunch tail catches the electrons at the beam head so that the bunch length decreases in the following drift space. When the high energy electron surpass the low energy electron and forms the bunch head, the energy chip flips and the beam begins to stretch again. Fig. 6b shows the velocity bunching process when the bunch compress is located 1 mm downstream the THz gun. With two 2 µJ THz pulse, the beam is compressed to a minimum r.m.s bunch length of 6 µm at z=5mm. The growth speed after the compressor is 1/30 of the speed before the compressor since the energy spread decrease from 0.8 keV to 0.1 keV.

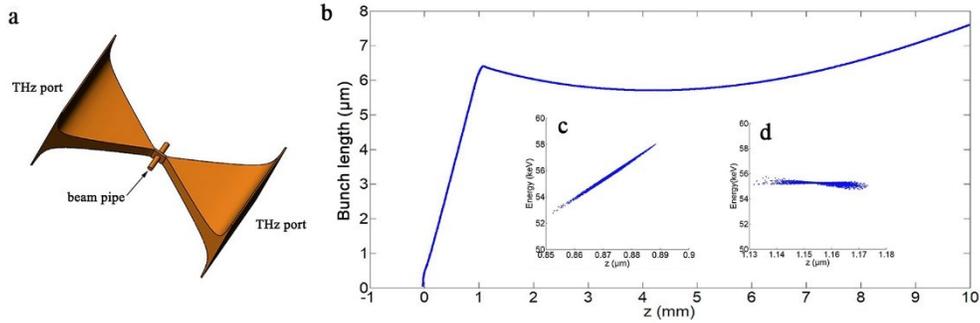

Fig. 6. The rectangular waveguide compressor. (a) the geometry of the compressor. (b) the r.m.s. bunch length along the beamline. Inset: the longitudinal phase space at the compressor entrance (c) and exit (d).

A uniform DLW works at the zero-crossing phase has been design to manipulate the longitudinal phase space of the electron beam produced by the injector. A uniform DLW benefits a longer interaction length than the rectangular waveguide compressor mentioned above. The relativistic electron beam form the injector moves faster than the nonrelativistic beam from the THz gun. The interaction slit of the rectangular compressor is 60 µm, correspond to a short interaction time, which requires much higher single cycle THz pulse energy to flip the energy chip. The interaction length of a uniform DLW depends on the designed phase velocity, group velocity and the THz pulse length, thus, a longer interaction length could be achieved by employing long uniform DLW powered by a relatively low energy few cycle THz pulse. Moreover, the transverse Lorentz force in the DLW tend to be zero when the electron velocity tend to the speed of light, which will contribute to a lower transvers emittance growth. Fig. 7a shows the geometry of the designed compressor, which is a 3 mm uniform quartz capillary with an inner diameter of 300 µm and a dielectric thickness of 92 µm. The longitudinal THz field is optimized to be fairly homogenous in the vacuum regime with only a 0.4% reduction from the vacuum axis to the dielectric wall. Fig. 7b shows the dispersion curve of the DLW, the phase velocity is $0.9894c$ with a corresponding group velocity of $0.03044c$. The DLW compressor is driven by a 10 ps, 0.45 THz THz pulse, dynamics simulation shows that the interaction length of the THz wave and the relativistic electron in the DLW compressor is 2.5 mm.

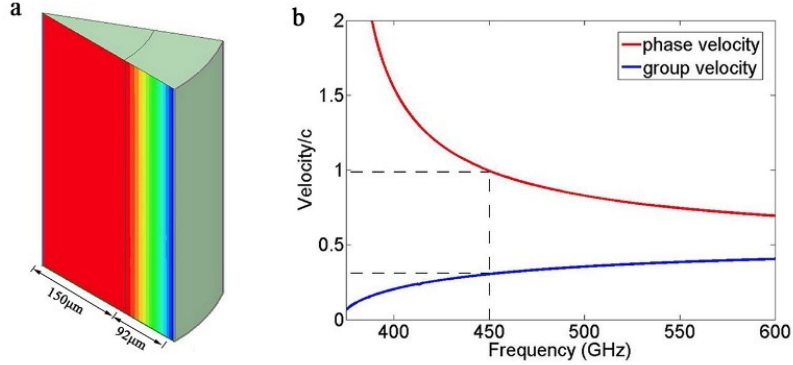

Fig. 7. The uniform DLW compressor. (a) the dimension and the field distribution of the DLW. (b) the dispersion curve of the DLW.

### 3.4 Permanent solenoid magnets

In order to avoid charge loss, the electron beam need to be properly focused before injected into the main linac or into the DLW compressor. Moreover, the transvers Lorentz force of the THz field and the space charge effects at the low energy regime will lead to strongly transverse emittance growth, which also requires carefully focusing to achieve proper emittance compensation. To address the above issue, we designed two types of PMSs by scaling the PMSs developed by REGAE at DESY [32]. PMSs provide stronger focusing strength, smaller weight and size than electromagnets without additional power source and cooling system. However, the focusing strength of a PMSs remains constant when the accelerator is working, the fringing field is also stronger than electromagnets which may lead to emittance growth. The designed PMSs consist of four circular magnets, each circular magnet is consists of 12 circular sectors made of neodymium (NdFeB) magnets, as shown in Fig. 8a. The remanence field of each sector is 1.47 T with opposite magnetic directions in the adjacent circular magnets. Fig. 8b shows the distribution of the magnetic field of the PMS. There are two types of PMS in the beamline, PMS1 encloses the low energy regime for emittance compensation, PMS2 focuses the beam to the DLW compressor and PMS3 focuses the beam at the end of the beamline. Fig. 8c shows the longitudinal magnetic field distribution of PMS1/PMS2 with a peak field of 1.05/0.95 T. PMS3 shares the same parameters of PMS2.

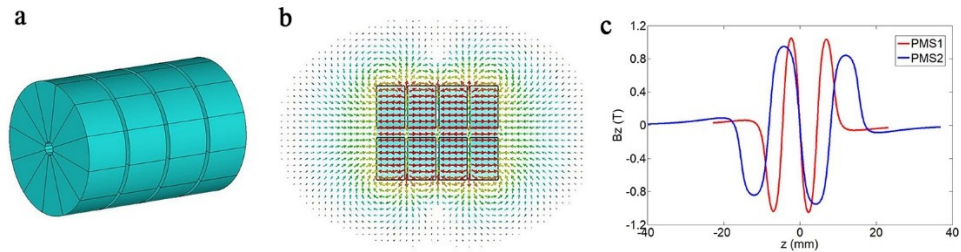

Fig. 8. The alternative permanent magnetic solenoid. (a) the geometry of the PMS. (b) the magnetic field distribution of the PMS. (c) the longitudinal magnetic field along the PMS vacuum axis.

## 4. Beam dynamics

### 4.1 Optimization of the transverse emittance of the injector

Owing to the space charge repulsion and the transverse Lorentz force of the DLW at the nonrelativistic regime, the transverse emittance continues to grow during the acceleration. In order to minimize the transverse emittance at the exit of the injector, we optimize the position and strength of the compensating magnet PMS1 and the main linac as well as the rectangular waveguide compressor via CST and ASTRA simulation. The optimized injector parameters are listed in table 1 and the beam parameter evolutions are shown in Fig. 9. The mean energy of the electron bunch is 3.00 MeV with an r.m.s energy spread of 25.94 keV, correspond to a relative energy spread of 0.86%. The beam is finely focused with a normalized transverse emittance of 0.073 π mm.mrad, which takes place at z =55.80 mm downstream the beamline. Owing to the space charge repulsion and fly speed divergence, the bunch length continues to grow after leaving the main linac, which requires further compression.

Table 1. The optimized parameters of the injector

| Parameter | Value | Unit |
| --- | --- | --- |
| The central position of rectangular compressor | 1.00 | mm |
| The peak electric field of rectangular compressor | 90.00 | MV/m |
| The position of the main linac entrance | 5.00 | mm |
| The peak electric field of the main linac | 285.00 | MV/m |
| The central position of PMS1 | 13.25 | mm |
| The peak magnetic field of PMS1 | 1.05 | T |

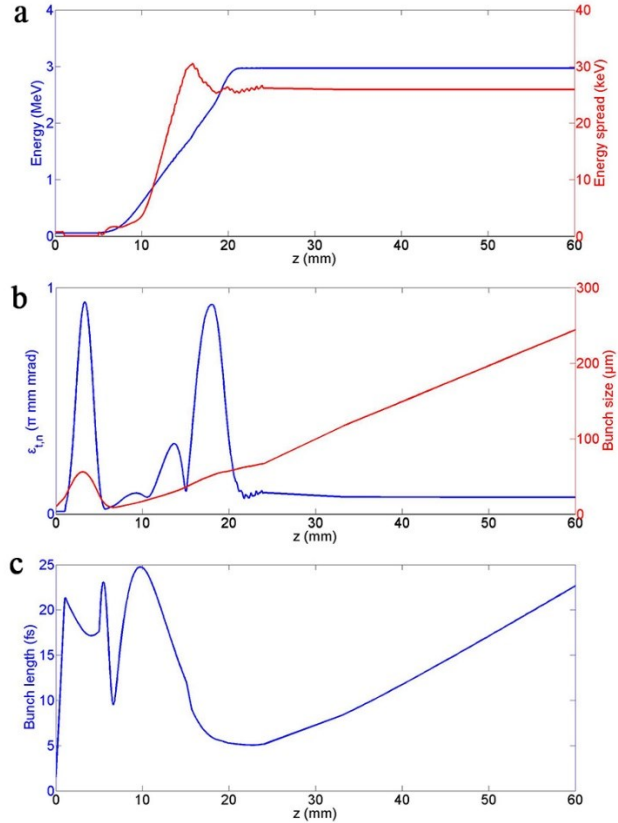

Fig. 9. The evolution of the phase space parameters of the injector. (a) the mean energy and r.m.s. energy spread, (b) the normalized transverse emittance and the r.m.s. bunch size, (c) the r.m.s. bunch length.

*4.2 Optimization of the bunch length and the energy spread*

The bunch length and the energy spread at the end of the beamline are controlled by the beam shaping line. The beam produced by the injector is imprinted with a negative energy chirp $h = d\delta/dt < 0$, where $\delta$ is the relative energy difference, that is, the electron energy at the bunch head is higher than that at the bunch tail. In the uniform DLW compressor, the bunch head is decelerated while the bunch tail is accelerated, a positive energy chirp is imprinted to the beam if the THz field is strong enough. The beam is then sending to a drift $L$ where the bunch tail catches the bunch head, resulting in a velocity bunching. Optimal compression is achieved when $hR_{56} = 1$, where $R_{56} \approx L/\gamma^2$ is the momentum compaction and γ is the Lorentz factor of the electron. The bunch length is related with the energy spread and usually could not obtain their minimum simultaneously. A minimum energy spread is preferred by many static applications such as static ultrafast electron diffraction and ultrafast electron microscopy, since a quasi-monochromatic beam will achieve a higher spatial resolution. However, most pump-probe experiment prefers a minimum bunch length so that it can probe the sample with a smaller time step. To address the above issues, we sweep the THz energy of the DLW compressor to find the corresponding configurations for a minimum bunch length or a minimum energy spread, as shown in Fig. 10a. In the range of 0~6 μJ, the DLW compressor significantly decreases the energy chip since the high energy bunch head is decelerated and the low energy bunch tail is accelerated. A minimum r.m.s. energy spread of 2.1 keV, correspond to a relative energy spread of 0.08% is achieved at 6 μJ THz energy with an r.m.s bunch lengh of 106.3 fs. The THz field is not strong enough to change the sign of the energy chirp and velocity bunch is not working. The bunch length decrease slowly since a low energy spread contributes to a low bunch stretching speed. In the range of 6~50 μJ, velocity bunching takes effect and the bunch is compressed quickly while the energy spread increases quickly. A minimum r.m.s. bunch length of 6.1 fs is achieved at 50 μJ THz energy with a relative energy spread of 1.69%. When the THz energy continuous to increase above 50 μJ, the bunch length slowly increase again since the beam is over compressed. Fig. 10b shows the normalized transverse emittance as a function of the compressor energy. The transverse Lorentz force focuses the beam at the compression phase which decrease the emittance at the low compression energy regime. When the velocity bunching takes place, the emittance of the compressed beam will increase owing to the Panofsky-Wenzel theory. The total effects of the transverse focusing and longitudinal compression makes the emittance drop at the beginning and grow after reaching a minimum of 0.078 π mm.mrad. The relative difference of the emittance at the range of 0~80μJ is less than 2.5%, which means that the DLW could effectively changing the energy spread and the bunch length while has little influence on the bunch emittance.

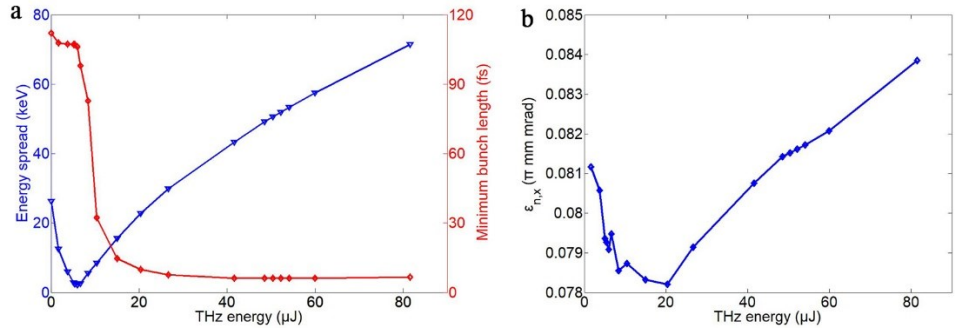

Fig. 10. The evolution of (a) the energy spread and minimum bunch length and (b) the normalized transverse emittance as a function of the THz energy of DLW compressor.

*4.3 Beam instability*

All the THz waves and the photo electrons are produced from the same seeding laser, the phase jitter induced by the optical path is negligible. However, the energy jitter of the power system still have a strong impact on the beam parameters. Fig. 11 shows the injection energy and the time of arrival (TOA) of the electron bunch at the entrance of the main linac under different THz energy. The energy jitter of the power system changes the injection energy and acceleration phase of the electron, and the compressor between the gun and the linac can significantly suppress the parameter jitter.

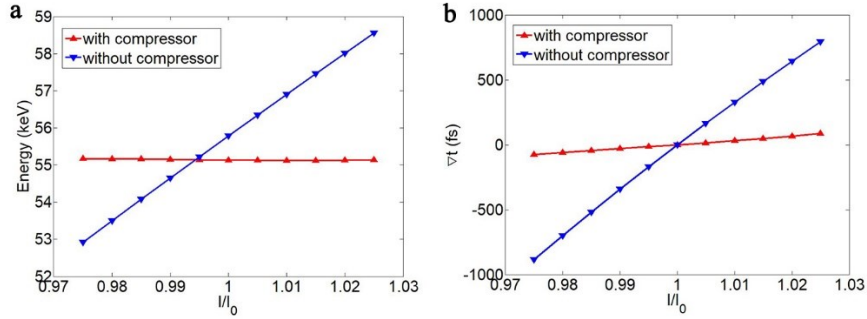

Fig. 11. The energy (a) and the divergence of TOA (b) at the entrance of the main linac under different THz energy.

The injection phase and the injection energy as well as the THz energy jitter of the main linac contribute to a beam parameter fluctuation of the injector as shown in Fig. 12. The beam parameters change more gently under a THz energy jitter regime of ±1.5% and significantly deviate from the design value if the THz energy jitter change above 1.5%.

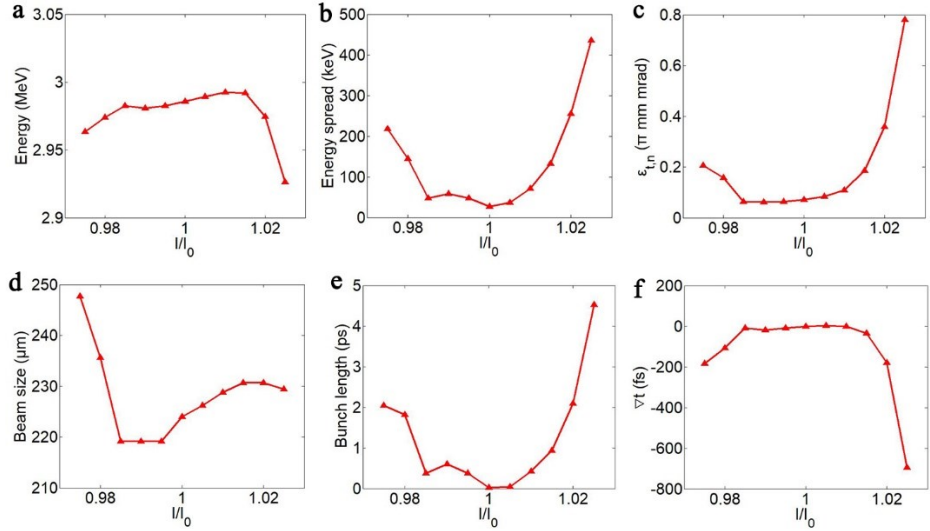

Fig. 12 The output parameters of the injector under different THz energies. (a) mean energy, (b) root-mean-square energy spread, (c) normalized transverse emittance, (d) root-mean-square bunch size, (e) root-mean-square bunch length and (f) the divergence of the TOA.

The fluctuation of the injector reduces the beam shaping results of the THz electron source as shown in Fig 13. The optimal bunch length and the minimum energy spread increase from the original value and stronger deviation appeals on the high energy side. The divergence of the TOA is strongly suppressed by the DLW compressor.

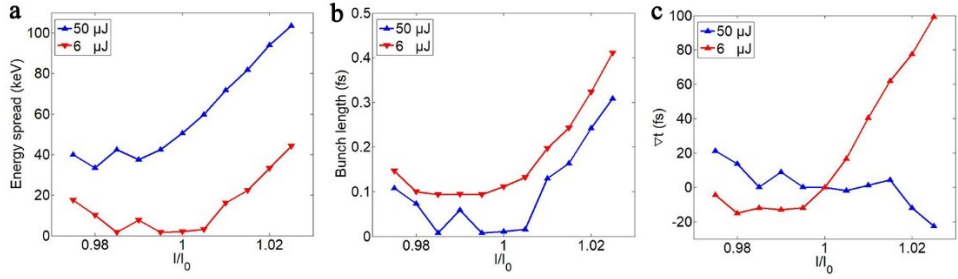

Fig. 13 The output parameters of the beam shaping line under different THz energies. (a) root-mean-square energy spread, (b) root-mean-square bunch length and (d) the divergence of the TOA.

## 5. UED based on the THz–driven electron source

To show the potential of the developed THz-driven electron source, we carry out UED simulation based on the produced electron beams using a similar code in ref [33]. The DLW compressor is working with 50 µJ THz energy to get a minimum relative energy spread of 0.08% since we are going to simulate a static UED pattern. A 312 nm thick aluminum foil is located at z=350 mm as the diffraction sample. The electron screen is located 0.5 m downstream the sample. Fig. 14a shows the simulated diffraction pattern. Fig. 11b shows the scattering intensity as a function of the momentum transfer $s = 4\pi sin(\theta/2)/\lambda$, where θ is the scattering angle and λ is the de Broglie's wavelength of the electrons. The diffraction circles up to the fourth order can be clearly distinguished from the simulated patter and the scattering intensity curve, showing that the developed compact all optical THz-driven electron source hold the great potential for achieving high quality MeV UED applications.

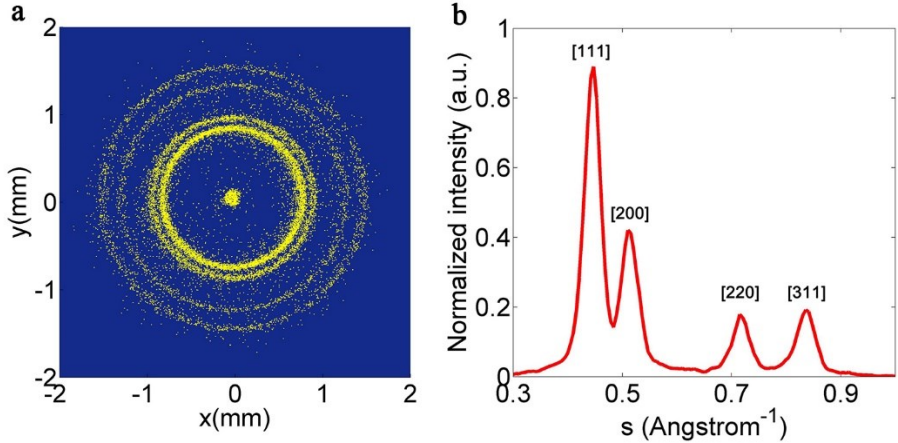

Fig. 14. The simulated diffraction pattern (a) and the corresponding scattering intensity (b).

## 6. Current status

The project of this compact all optical THz-driven electron source was firstly proposed in 2021 [24] based on our previous works on cascaded THz-driven acceleration [16], THz-based beam diagnostics [25], frequency tunable single-cycle or multi-cycle optical THz sources [26] and THz streaking method. We have finished the conceptual and technical design of the project now, and we are going to work on the preliminary experimental demonstration of the key components recently. A prototype THz-driven electron gun has been fabricated and is now under testing. Fig. 15 shows the assembly of the THz gun and the experimental setup of the beam test. Fabrication of the tapered DLW and the PMSs is on the schedule and the Joule-level multi-cycle THz source is still need extensive efforts, more results will be reported in future works.

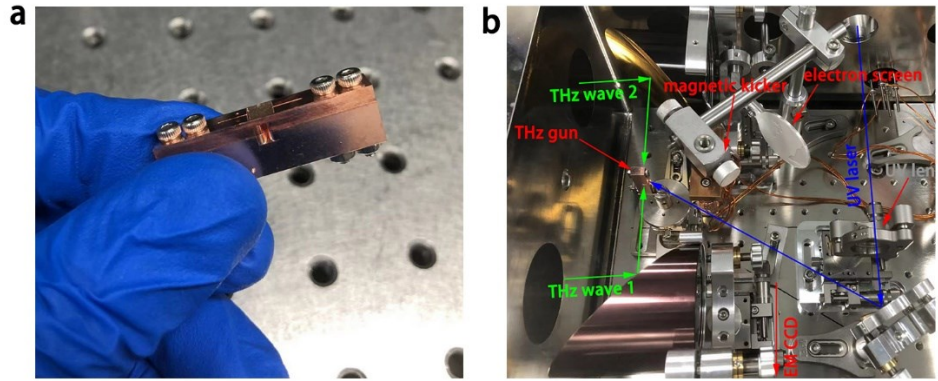

Fig. 15. (a) the assembly of the THz gun and (b) the experimental setup of the beam test.

## 7. Conclusion

We have proposed a compact all optical THz-driven electron source at Tsinghua University. The physical design of the THz-driven electron gun, the tapered DLWs linac, the focusing PMS and the THz-driven bunch compressors have been included. Dynamics optimizations of the beam line have produced a 3 MeV electron beams with a normalized transverse emittance of 0.079 π.mm.mrad. A minimum relative energy spread of 0.04% or a minimum r.m.s. bunch length of 6.1 fs can be achieved by adjusting the beam shaping line. Sensitivity analysis shows that the THz-driven electron source can effectively work under a 1.5% energy jitter of the THz power system. High quality diffraction pattern has been simulated showing the great potential of the design THz-driven electron source. This compact all optical THz electron source reveals the full capability of THz-driven acceleration technology, providing a feasible technique approach for build a real THz-driven accelerator, which holds great potential for high quality electron related scientific researches. We are now working on the preliminary experimental demonstration of prototype THz-driven electron gun and other beam line components. More works will be reported in future works.

**Funding.** National Natural Science Foundation of China (NSFC Grant No. 12035010).

**Acknowledgments.** We thank Prof. Yutong Li, Prof. Jinglong Ma, Ph.D student Baolong Zhang from the Institute of Physics, Chinese Academy of Sciences, and Prof. Xiaojun Wu from Beihang University for their help in building the experimental platform for the beam test of the THz-driven electron gun.

**Disclosures.** The authors declare no conflicts of interest.